\documentclass[aps,prb,twocolumn,floatfix]{revtex4}
\usepackage{graphicx}
\begin{document}
\title{Hofstadter butterfly for a finite correlated system}
\author{Katarzyna Czajka, Anna Gorczyca, Maciej M.  Ma{\'s}ka, and Marcin Mierzejewski} 
\affiliation{
Department of Theoretical Physics, Institute of Physics, 
University of Silesia, 40-007 Katowice,
Poland}

\begin{abstract}
We investigate a finite two--dimensional system in the presence of external 
magnetic field. We discuss how the energy spectrum depends on the system 
size, boundary conditions and Coulomb repulsion. On one hand, using 
these results we present the field dependence of the transport properties 
of a nanosystem. In particular, we demonstrate that these properties depend 
on whether the system consists of even or odd number of sites.  On the other 
hand, on the basis of exact results obtained for a finite system we 
investigate whether the Hofstadter butterfly is robust against strong 
electronic correlations. We show that for sufficiently strong Coulomb 
repulsion the Hubbard gap decreases when the magnetic field increases.
\end{abstract}
\pacs \ 73.22.-f, 73.63.-b, 71.70.Di
\maketitle
\section{Introduction}

The problem of electrons moving in a periodic potential under the influence
of an external magnetic field has been investigated since the beginning of 
quantum mechanics. Despite the seeming simplicity of the problem, many its
aspects still remain unresolved. Even in the absence of electronic correlations
solutions are known only in limiting cases. In particular, two dimensional (2D)
electron gas under the influence of a periodic potential and a perpendicular
magnetic field can be described in two limits, one of a weak and the other
of a strong periodic potential. In the former case the applied magnetic field
is the main factor that determines the behavior of electrons. As a result
the electronic wave functions are Landau level--like, with the degeneracy
lifted by the periodic potential. If the potential is modulated in one
dimension, the width of the resulting ``Landau bands'' oscillates with
the magnetic field as a consequence of commensurability between the 
cyclotron diameter and the period of potential modulation. It leads to
oscillations in magnetoresistance, known as the Weiss oscillations.\cite{Weiss}
If the potential is modulated in two dimensions, ``minigaps'' open in the 
``Landau bands'', and the energy spectrum plotted versus the applied field 
composes the famous Hofstadter butterfly.\cite{langbein,Hofstadter} It is 
interesting, that the same spectrum occurs in a complementary limit, when 
the lattice potential is very strong, and the electronic wave functions are 
Bloch--like, modified by the magnetic field.\cite{langbein}

The simplest model for the case, when an applied field and a lattice
potential are present simultaneously, is commonly referred to as the Hofstadter
or Azbel--Hofstadter model.\cite{Hofstadter,Azbel} The corresponding
Hamiltonian
describes electrons on a two--dimensional square lattice with
nearest--neighbor hopping in a perpendicular uniform magnetic
field. The Schr\"odinger equation takes the form of a one--dimensional
difference equation, known as the Harper equation (or the almost
Mathieu equation).\cite{Harper,Hofstadter,Rauh}
It is also a model for a one--dimensional electronic system in two
incommensurate periodic potentials. The Harper equation also has
links to many other areas of interest, e.g., the quantum Hall effect,\cite{Albrecht}
quasicrystals, localization--delocalization phenomena,\cite{bellissard,guarneri} 
the noncommutative geometry,\cite{bellissard1}
the renormalization group,\cite{thouless1,wilkinson} the theory of
fractals, the number theory, and the functional analysis.\cite{last1}
It is also useful in determining the upper critical field
\cite{mmmm2,mmmm1,mmmm3,dmm}
and the pseudogap closing field \cite{mmmm4} in high--temperature
superconductors.

The unusual structure of the Hofstadter butterfly is a characteristic feature of 2D systems.
Similarly to the case of Landau levels the dimensionality  of the system 
is of crucial importance for the
Hofstadter butterfly. Movement of electrons along 
the external magnetic field would be responsible for broadening
of the Hofstadter bands. They eventually may overlap and, in this way, smear out the original 
fractal structure of the energy spectrum. On the other hand it is known,
that electronic properties of low dimensional systems may be completely
changed by the presence of electronic correlations. In particular, 
it is well known for 
1D systems that perturbation theory breaks down and arbitrarily weak 
on-site Coulomb repulsion qualitatively changes the whole excitation   
spectrum from the Fermi to Luttinger liquid type. 
The role of Coulomb interaction in the case of 2D lattice
has intensively been investigated in connection with high temperature
superconductors. Although a complete description of the correlated
2D system is still missing, 
it became obvious that mean-field approaches 
are invalid even for moderate values of Coulomb repulsion. Therefore,
a question arises, whether the fine structure of  the Hofstadter butterfly 
is robust  against the presence of electronic correlations. 
This problem has previously been investigated 
on a mean--field level.\cite{gg,doh}
In particular, the analysis presented in Ref. \onlinecite{doh} suggests that also in the presence 
of electronic correlations the energy levels should form the Hofstadter butterfly 
with additional energy gap in the middle of the energy spectrum. 
However, the mentioned above limitations of the mean-field results
clearly show, that these results do not represent a conclusive solution. 
In this paper we address this problem with
the help of a method that is particularly suited for investigations
of the low dimensional correlated systems, i.e, exact diagonalization
of finite clusters. Although in this method short-range electronic correlations
are exactly taken into account, finite size effects will seriously affect
the energy spectrum. These modifications can be of special importance, when the system 
is under influence of magnetic field.\cite{anal1,anal2}
In order to separate the correlations and size--induced 
effects 
we start our investigations with a finite uncorrelated system and discuss
both fixed (fbc) and periodic boundary conditions (pbc).
Apart from the discussion of
the Hofstadter butterfly these results may be applicable
to investigations of nanosystems in the presence of magnetic field, where
fixed boundary conditions are more appropriate than the periodic ones.

\begin{figure}
\includegraphics[width=8.2cm]{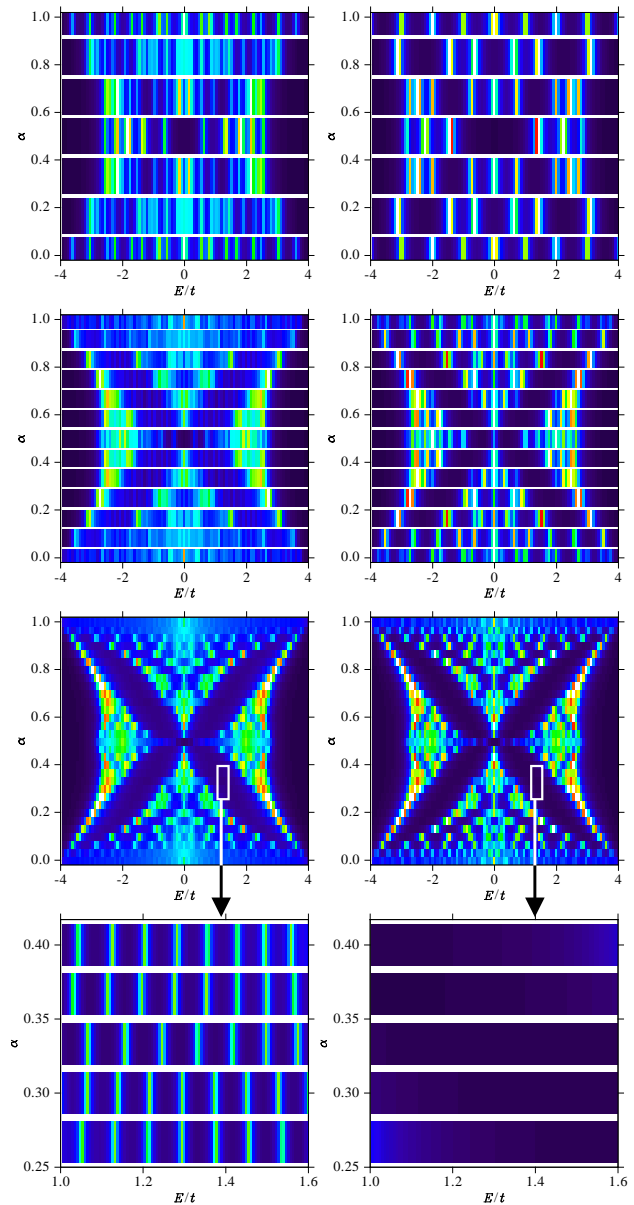}
\caption{(Color online). Density of states, $\rho(E)$, as a function of the magnetic field
($\alpha$) calculated with fixed (left column) and periodic (right column) boundary
conditions [see Eq. (\ref{rhodef})]. 
The brighter colors correspond to larger values of the density of states.
The first, second and third row shows results obtained for a
$6 \times 6$, $12 \times 12$ and $30 \times 30$ cluster, respectively.
The broadening of the one--particle levels $\eta=0.05t$ has been assumed. 
In the last row we magnify the regions marked above. In order to demonstrate
the presence of the edge states, the broadening has been reduced to
$\eta=0.005t$.
}
\label{result}
\end{figure}
\section{Finite--size effects}
For the sake of completeness, we start with a brief derivation of the Harper equation for 
the gauge $ {\bf A}=B\left(-ay,(1-a)x,0 \right) $. Here, the parameter $a \in (0,1)$ allows one
to distinguish between the Landau ($a=0$) and symmetric gauges ($a=1/2$).
The 2D square lattice in the presence of external, perpendicular magnetic field can be described by 
the tight--binding Hamiltonian:
\begin{eqnarray} 
H=t\sum_{x,y,\sigma} \left({\mathrm {e}}^{  i \phi_x y}c_{(x,y)\sigma}^{\dagger}
c_{(x-1,y)\sigma}+{\mathrm {e}}^{-  i \phi_x y}c_{(x,y)\sigma}^{\dagger} 
c_{(x+1,y)\sigma}\right .&&\nonumber \\ 
+ \left.{\mathrm {e}}^{-  i \phi_y x} c_{(x,y)\sigma}^{\dagger}c_{(x,y-1)\sigma}
+{\mathrm {e}}^{  i  \phi_y x}c_{(x,y)\sigma}^{\dagger}c_{(x,y+1)\sigma}\right), 
\nonumber \\
\label{hamu0}
\end{eqnarray}
where $c^{\dagger}_{(x,y)\sigma}$ creates an electron with spin $\sigma$ at the site $(x,y)$ and
$t$ is the nearest--neighbor hopping integral in the absence of magnetic field.
$\phi_x=2 \pi \alpha a $, $\phi_y=2 \pi \alpha (1-a) $, 
$\alpha= \Phi/\Phi_{0} $, where $\Phi$ is the magnetic flux through the lattice cell and $\Phi_0$ 
is the flux quantum. 
In order to determine eigenfunctions of this Hamiltonian, formally
one should solve a 2D eigenproblem.  
However, in the case of the Landau gauge ($a=0$), the hopping integrals
in Eq. (\ref{hamu0}) depend solely on the $x$ coordinate. Because of the translational invariance
along the $y$ axis, eigenfunctions exhibit a plane--wave behavior in this direction 
[$\exp(i k y)$].  This argumentation can be extended to a more general gauge.
Such an extension requires an appropriate shift of the momentum 
$k$.
For a $N_x \times N_y$ lattice with pbc, we introduce  fermionic operators  
$c_{x,k,\sigma}^{\dagger}$ defined by 
\begin{equation}
c_{(x,y)\sigma}^{\dagger}=
\frac{1}{\sqrt{N_{y}}}\sum_{k}{\mathrm{e}}^{-iy \left(k+\phi_x x\right) }c_{x,k,\sigma}^{\dagger}.
\end{equation}
The Hamiltonian (1) can then be written:
\begin{eqnarray}
H&=&t \sum_{x,k,\sigma}  \left(c_{x,k,\sigma}^{\dagger}c_{x+1,k,\sigma}+
 c_{x,k,\sigma}^{\dagger}c_{x-1,k,\sigma} 
\right. \nonumber \\
& &\left. \qquad \mbox{}+
2{\mathrm{cos}}(k-  2 \pi \alpha x)c_{x,k,\sigma}^{\dagger}c_{x,k,\sigma}
\right).
\label{ham3D}
\end{eqnarray}
The resulting Hamiltonian is diagonal in the quantum numbers $k$ and 
tridiagonal in the $x$-coordinates. It means that the applied 1D 
transformation to the momentum space, allows one to reduce the original 2D 
eigenproblem [Eq. (\ref{hamu0})] to a 1D one. In the following we study 
whether such a reduction is possible also for a finite system with fixed 
boundary conditions. The relevant wave--function in this case is no longer the 
plane-wave, since it must vanish at the system edges. It is easy to check, 
that the Hamiltonian of a 1D chain with fbc can be diagonalized with the help 
of the transformation
$
c_{y,\sigma}^{\dagger} \sim \sum_{k}{\mathrm{sin}}(ky)c_{k,\sigma}^{\dagger}
$.
In the case of a 2D lattice in the presence of the external magnetic field, 
an appropriate shift of the wave--vectors allows one to cancel the
Peierls phase factors for the hopping along the $y$-axis: 
\begin{displaymath}
c_{(x,y)\sigma}^{\dagger}=\sqrt{\frac{2}{N_{y}+1}}\sum_{k}{\mathrm{sin}}(ky)c_{x,k,\sigma}^{\dagger}
{\mathrm e}^{2 \pi i \alpha xy},
\end{displaymath}
where the wave--vectors 
$$
k =  \frac{\pi}{N_{y}+1},\;\frac{2\pi }{N_{y}+1},\;\dots,\;\frac{N_{y}\pi}{N_{y}+1}.
$$
In the above equation we have assumed $a=1$. Generalization to arbitrary value of
$a$ is straightforward.

The orthogonality relation
$$
\sum_{y}{\mathrm{sin}}(ky){\mathrm{sin}}(py)=\frac{N_{y}+1}{2}\delta_{kp}
$$
allows one to carry out the  inverse transformation. Then, the Hamiltonian takes the form
\begin{eqnarray*}
H=t \sum_{k,\sigma} \Bigg(
\sum_{x=1}^{N_{x}-1}\sum_{y=1}^{N_{y}}
\frac{2}{N_{y}+1}\sum_{p}{\mathrm{sin}}(ky) {\mathrm{sin}}(py)\times && \\
 \Big({\mathrm{e}}^{- 2\pi i \alpha y}c_{x,k,\sigma}^{\dagger}c_{x+1,p,\sigma}+h.c. \Big)+
\sum_{x=1}^{N_{x}}2{\mathrm{cos}}(k)c_{x,k,\sigma}^{\dagger}c_{x,k,\sigma} \Bigg). &&
\end{eqnarray*}  
Only the term responsible for the hopping along the $y$-axis is diagonal in $k$,
whereas the remaining hopping term is generally not. Therefore, contrary to the case of infinite 
lattice with pbc, the Hamiltonian cannot be reduced to a form that is diagonal
in wave-vectors and tridiagonal in real space coordinates. The only exceptions occur 
for $\alpha=0$ and  $\alpha=1/2$. The first case is trivial. 
In the latter case ($\alpha=1/2$), an additional transformation
\begin{displaymath}
c_{x,k,\sigma} \rightarrow \frac{1}{2}c_{x,k,\sigma}\left(1+(-1)^{x}\right)
-\frac{1}{2}c_{x,\pi-k,\sigma}\left(1-(-1)^{x}\right),
\end{displaymath}
leads to the Hamiltonian in the form given by Eq. (\ref{ham3D}). 
However, since $k \in (-\pi,\pi)$ for pbc and 
$k \in (0,\pi)$ for fbc, also in this case the energy spectrum
depends on the boundary conditions. 
Consequently, degeneracy of the energy levels is lower for fbc than for pbc. 
An additional difference, that immediately 
follows from the analytical calculations, is related to
 the density of states $\rho(\epsilon)$ for $\alpha=1/2$. In 
the case of infinite system with pbc $\rho(0)=0$, whereas for 
a finite lattice with fbc there exists an eigenvalue
$\epsilon=0$, provided the system consists of odd number of sites.
The proof of this statement is 
straightforward and, therefore, we omit the details. The difference between systems with 
even and odd number of sites will be discussed in more details in connection with the 
transport properties. 
In Fig. 1 we compare the density of states obtained for various 
boundary conditions and lattice sizes. It has been calculated using the
standard formula: 
\begin{equation}
\rho(E)=-\frac{1}{\pi N_x N_y} \sum_n \Im \frac{1}{E-\varepsilon_n+i \eta},
\label{rhodef}
\end{equation}
where $\varepsilon_n$ is the energy of the $n$-th one--particle eigenstate.

In the presence of magnetic field the translation group of the lattice does not represent
a symmetry group of the Hamiltonian and one can discuss periodicity only 
with respect to the magnetic translation group.\cite{zak,floratos} Consequently,
for a finite system one can apply the pbc only for some particular values of $\alpha$,
which are determined by the system size.
On the other hand, for fbc the energy levels can be calculated for arbitrary magnetic field, 
as it was demonstrated
in Ref. \onlinecite{anal2}. However, in order to compare directly results obtained for fbc and pbc,
in both the cases we have used only these values of magnetic field, which are allowed for pbc.
In the case of fbc there exist edge states, which are responsible for additional levels 
inside the energy gaps in the Hofstadter butterfly.\cite{levels_in_gaps}
They are clearly visible in Fig. 1 in the density of states obtained 
for a $6\times 6$ system.
The relative 
contribution of these states to $\rho(\epsilon)$ decreases with the system size.  
Therefore, they are much less visible in the density of states obtained 
for larger systems 
and become unimportant for an infinite lattice. 
However, for a finite system, they may qualitatively change the 
well known structure of the Hofstadter butterfly, obtained from the Harper equation.

\subsection{Transport properties}

The above discussion can directly be applied to the investigations
of transport properties of nanosystems. 
The field--induced modifications of the energy spectra lead
 to strong changes of the transport current, at
least in the low voltage regime.
We use the formalism 
of nonequilibrium Green functions to
analyze these effects in a nanosystem coupled to leads. 
The coupling is described by:
\begin{equation}
H_{\rm nano-el}=\sum_{{\bf k},x,y,\alpha,\sigma} \left(
g_{{\bf k} ,x,y,\alpha} d^\dagger_{{\bf k},\sigma,\alpha }c_{\left(x,y\right)\sigma}+{\rm H.c.}\right),
\end{equation}
where $d^\dagger_{{\bf k},\sigma,\alpha} $ creates an electron with
momentum ${\bf k}$ and spin $\sigma$ in the electrode $\alpha$. Here,  $\alpha \in$ \{L,R\} indicates the left or
right electrode.
We assume a simple model in which the leads are described by a two--dimensional (2D) lattice gas and the
hopping between the leads and nanosystem is possible only perpendicularly to the edge
of the nanosystem. $g_{{\bf k},x,y,\alpha}$ is nonzero  
only for sites $(x,y)$ which are located at the edge $\alpha$ of the nanosystem.
The details of calculations can be found in Refs. 
\onlinecite{my} and \onlinecite{bulka}. 
Figs. 2 and 3 show the field dependence 
of the transport current for various size of the nanosystem.

\begin{figure}
\includegraphics[width=8cm]{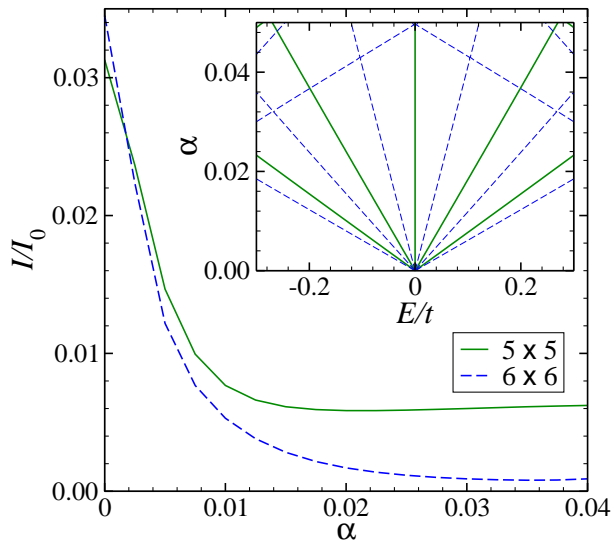}
\caption{(Color online) Field dependence of the transport current through the $5 \times 5$ 
and $6 \times 6$ nanosystems coupled to macroscopic leads for the applied voltage
$eV=0.01t$.  $I_0=2 e t/h$, where $e$ is the elementary charge and $h$ is the Planck constant. See Fig. 5
in Ref. \onlinecite{my} for parameters and the details of the coupling between the nanosystem and the electrodes.
The  inset shows the field dependence of the one--particle eigenenergies located in the vicinity of $E=0$ obtained for an isolated
nanosystem with fbc.
}
\label{result1}
\end{figure}
\begin{figure}
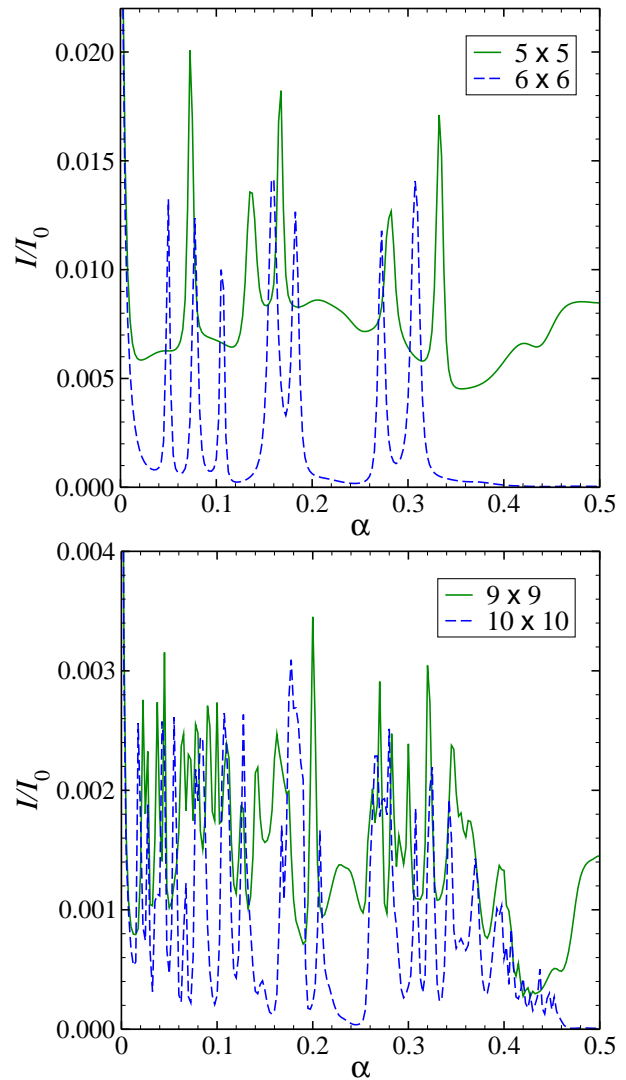

\includegraphics[width=8cm]{fig3a.eps}
\includegraphics[width=8cm]{fig3b.eps}
\caption{(Color online) 
Upper panel shows the field dependence of the transport current through the $5 \times 5$ 
and $6 \times 6$ nanosystems for the applied voltage $eV=0.01t$. 
Lower panel shows results obtained  for $9 \times 9$ 
and $10 \times 10$ nanosystems with  $eV=0.002t$. The remaining parameters are the
same as in Fig. 2. 
}
\label{result2}
\end{figure}
Two main features arise from the
presented results: (i) strong magnetoresistance in a weak field regime and (ii) even--odd parity effect.
Both these features occur for a low voltage only. The first effect is shown in Fig. 2. 
It can easily be understood on the basis of the field dependence of the one--particle energies that are 
close to the Fermi energy, as presented in the inset in Fig. 2. One can see, that for low voltage the 
number of states which participate in the transport decreases when the magnetic field increases. 
This effect is due to a field--induced splitting of a strongly degenerated level at zero energy. 
This degeneracy, in turn, is a remnant of the van Hove singularity, that is a typical feature of an 
infinite 2D lattice. There is, however, a significant difference between systems with even and odd number 
of lattice sites. In the former case, strong magnetic field can completely remove states from the 
vicinity of the Fermi energy, what results in vanishing of the current. On the other hand, 
if there is an 
odd number of lattice sites, one state is always located at zero energy.   
As a result a finite conductivity occurs for arbitrary magnetic field.
This parity effect originates from the facts that the energy
spectrum is symmetric with respect to the zero energy and the number of energy levels is equal to the number of
lattice sites.
The difference between systems consisting of even and odd number of 
sites is pronounced for $\alpha=1/2$, what can be seen in Fig. 3. The even--odd parity effect
is a well known feature of persistent currents in mesoscopic rings, where it occurs due to its nontrivial
first homotopy group.\cite{ring} 
Here, we have demonstrated that a similar effect may occur also in a nanosystem with a trivial 
topology: in an isolated nanosystem as well as in a system coupled 
to a macroscopic leads. In the first case it shows up in the energy spectrum, whereas in the
latter case it is visible also in the transport properties. 
An additional similarity between rings and the systems under investigation concerns the fact, that the parity effect
occurs only in small systems and disappears in the thermodynamic limit.

\section{The role of correlations}
\begin{figure*}[t]
\includegraphics[width=16cm]{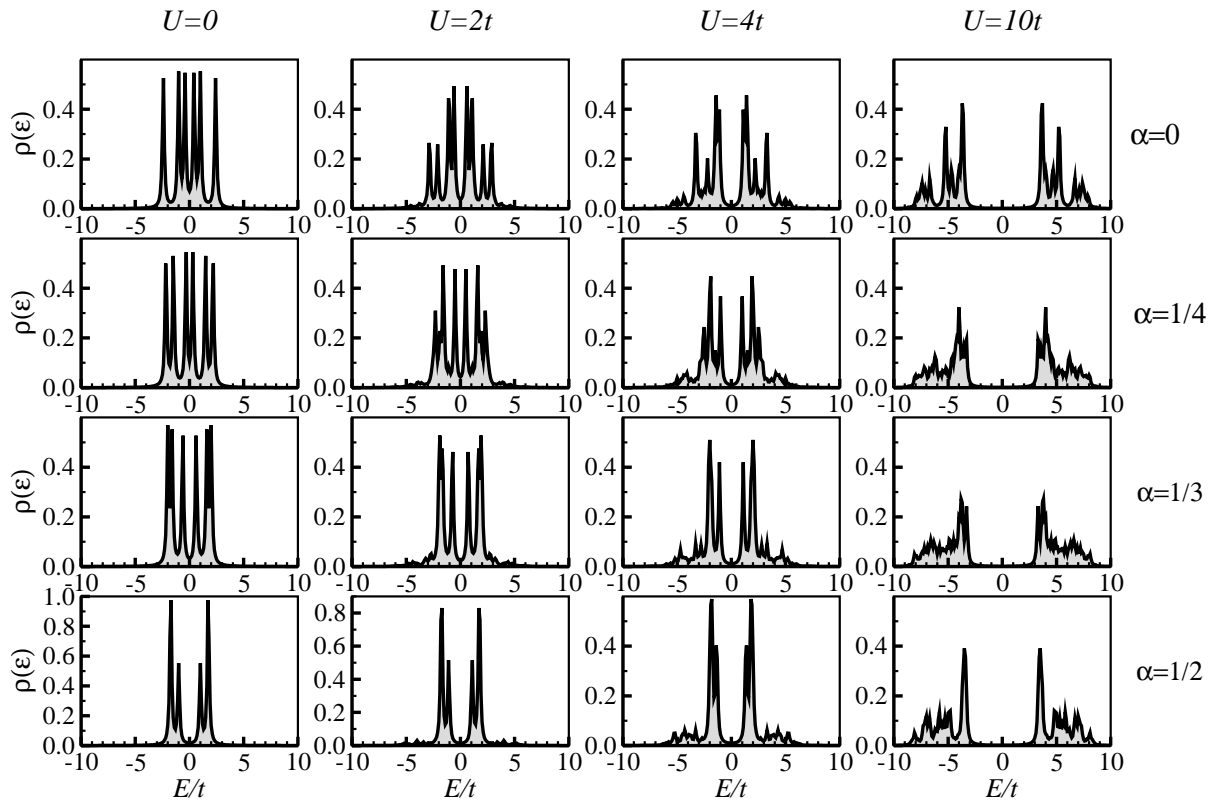}
\caption{Density of states obtained for a  $2 \times 3$ cluster with fbc for various values of the Coulomb
repulsion and the magnetic field.}
\label{result3}
\end{figure*}

After establishing the role of boundary conditions 
and its significance for the properties of nanosystems, 
we switch to the main question, whether the Hofstadter energy
spectrum is robust against the presence of strong electronic correlations. 
Both the external magnetic field and electronic correlations
give rise to opening of the energy gaps in the density of states. One could expect 
that these mechanism should independently contribute to the opening of these gaps.
In the following, we demonstrate that this intuitive statement is wrong and, actually, external
magnetic field reduces the Hubbard energy gap. In order to investigate this problem, we consider
the 2D Hubbard model in the presence of magnetic field:
\begin{equation}
H_{{\mathrm Hubb}}=H_{\mathrm kin}+U \sum_{x,y} n_{\left(x,y\right) \uparrow}  n_{\left(x,y\right) \downarrow},
\end{equation}
where the kinetic term is given by Eq. (\ref{hamu0}) and the electron number operator 
$ n_{\left(x,y\right)\sigma}=c_{(x,y)\sigma}^{\dagger}c_{(x,y)\sigma}$. 
In the following we investigate the half--filled case, when the 
Hubbard gap opens at the Fermi level, driving the system from a metallic state 
to an insulating one.

This Hamiltonian has exactly been diagonalized  mostly by means of
the Lancz\"os method. It is one of the most effective computational
tools for searching for the ground state and some low laying excited
states of a finite system.
We start with a system sufficiently small to allow one
to determine the whole energy spectrum. 
In the case of finite system calculations pbc can be applied 
only for specific values of the magnetic field, which depend on the cluster size.
Therefore, in the following we use the fbc. 
Fig. 4 shows how the density of states depends on the
magnetic field and the magnitude of the Coulomb repulsion.  One can see, that for weak to moderate Coulomb
repulsion its influence on the density of states strongly depends on the applied magnetic field.
In the absence of magnetic field ($\alpha=0$) even relatively weak electronic correlations ($U=2t$) strongly
modify the energy spectrum. On the other hand, for strong magnetic field
the density of states becomes robust against the Coulomb correlations. 
In particular, for $\alpha=1/2$ densities
of states obtained for $U=0$ and $U=2t$ hardly differ from each other. 
This result already suggests that Coulomb
interaction modifies various parts of the Hofstadter butterfly in a different way. It remains in 
contradiction to the results obtained in the mean-field analysis.\cite{doh} In the latter
approach the Coulomb repulsion opens an almost field independent gap in the middle 
of the Hofstadter butterfly. Additionally, the one-particle energies remain in the range
$(-4t,4t)$, at least for moderate $U$, i.e., the electronic correlations do not change the
spectrum width. It has also been reported in Ref. \onlinecite{doh} that for
stronger Coulomb repulsion the only modifications of the Hofstadter butterfly concern
the larger band gaps and narrower band widths. Contrary to the mean--field results, exact 
diagonalization method
indicates that electronic correlations lead to a substantial  modification of the density of states.
First, there are excitation beyond the range $(-4t,4t)$. When $U$ increases
the one-particle excitations are replaced by the collective ones with a finite live-time.
Instead of the $\delta$--like peaks in the density of states we obtained much larger number of
wider peaks, which eventually may overlap.
Therefore, we expect that electronic correlations smear out the fine structure 
of the Hofstadter butterfly.

For the application purposes, the most important property concerns the density of states in the
vicinity of Fermi level. Therefore, we restrict the following study only to
the ground state and the lowest excited states.  Especially, we analyze field dependence of the Hubbard gap 
$\Delta(U,\alpha)$ that opens in the middle of the density of states.
In Fig 5. we present a reduced gap $\Delta^*(U,\alpha)=\Delta(U,\alpha)-\Delta(U,0)$, 
 whereas $\Delta(U,0)$ is presented in the inset in this figure. 

\begin{figure}
\includegraphics[width=8.5cm]{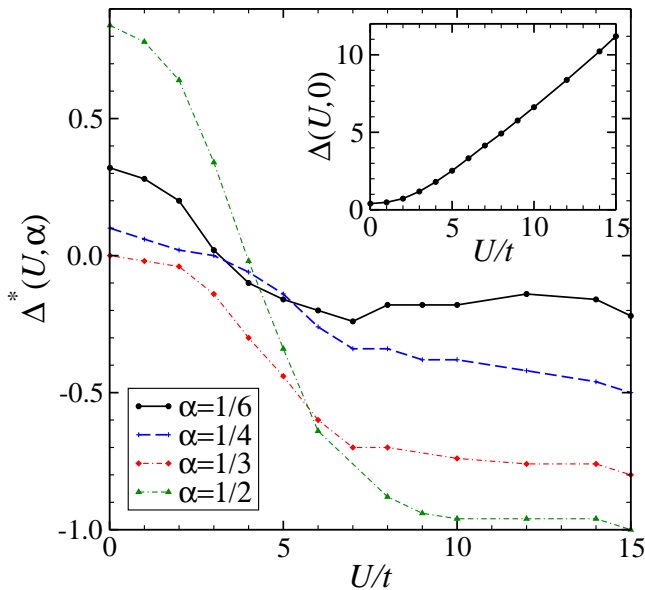}
\caption{(Color online)
Magnitude of the reduced Hubbard gap ($\Delta^*(U,\alpha) $)  
as a function of $U$ for various magnetic fields. These results 
have been obtained for a $12$--site cluster with fbc.}
\label{result4}
\end{figure}

As one could expected, $\Delta$ increases with $U$ for arbitrary magnetic field.
However, the negative slope of  $\Delta^*$ as a function of $U$ clearly indicates that this increase is
smaller when the magnetic field is switched on. Finally for sufficiently strong Coulomb repulsion, the
Hubbard gap seems to be a monotonically decreasing function of magnetic field for $\alpha \in (0,1/2)$.  
Since the Hamiltonian is invariant under the transformation $\alpha \rightarrow 1-\alpha$ 
the gap monotonically increases with $\alpha$, for $\alpha \in (1/2,1)$. 
This monotonic behavior is an unexpected result that
strongly contrasts with the uncorrelated case, where 
the gap changes irregularly (discontinuously) with magnetic field, as
can be inferred from Fig. 1.

 The Hubbard model has already been applied for investigations of nanosystems, e.g.,  molecular wires\cite{bulka}
and quantum dot arrays (see Ref. \onlinecite{d2} for a detailed discussion and estimation of
the relevant model parameters).
The latter case is of particular interest since the size of the elementary
cell can be adjusted in such a way, that the flux through the cell can be of the order of the flux quantum and,
then, the structure of the Hofstadter butterfly may be visible. Unfortunately, the present
approach is probably oversimplified for a quantitative description of electronic correlation 
in the quantum dot arrays, where one should account for a large
number of states per dot. In spite of this, the considered Hubbard Hamiltonian allows for at most
two electrons per site. Additionally, we have restricted our considerations only to the on--site repulsion $U$, what
corresponds to the intradot interaction. In a more realistic approach, an extended multiband Hubbard model with
intersite interaction should be used, 
but such extensions are presently beyond the reach of the exact diagonalization techniques.
\section{summary}

The structure of the Hofstadter butterfly arises due to coexistence of the periodic
potential and the perpendicular magnetic field in the 2D electron gas. It is well known that
the presence of Coulomb interaction in low dimensional systems is responsible for strong
modification of its electronic properties. Motivated by these facts, we have investigated
how the butterfly structure is affected by the correlations. 
In contrast to the mean--field results we have demonstrated that  Coulomb correlations are 
responsible for broadening of the quasiparticle levels. As a result some of the Hofstadter 
bands overlap, and the fractal butterfly structure smears out. We have shown that for 
sufficiently strong repulsion the Hubbard gap monotonically decreases with the external 
magnetic field for $\alpha \in (0,1/2)$. This result strongly contrasts with the irregular 
behavior of the density of states in the uncorrelated case. Unfortunately, beyond the 
mean--field level, results can be obtained only for relatively small systems. 
We expect, however, that similar changes occur also in
much larger systems. 
On the other hand, some of the results can directly be applied to the investigations
of nanosystem, e.g. quantum dot arrays.
The finite--size effects seriously modify
field dependence of its transport properties. Similarly to the
persistent currents in nano-- and mesoscopic rings, also the transport currents 
are different in systems consisting of odd and even  number of sites.

\acknowledgments
This work has been supported by
the Polish Ministry of Education and Science under Grant No. 1~P03B~071~30.

\end{document}